 \documentclass[final,1p,times]{elsarticle}

\usepackage{amsmath,amssymb}
\usepackage{subfigure,graphicx}
\usepackage{amsmath}
\usepackage{amsfonts}
\usepackage{amssymb}

\journal{Nuclear Physics B}

\begin{document}

\begin{frontmatter}
\title{$(Z\alpha)^{4}$ order of the polarization operator in Coulomb field at low energy}
\author[BINP]{G.G.Kirilin}
\ead{G.G.Kirilin@inp.nsk.su}
\author[BINP,NSU]{R.N. Lee\fnref{DFG}}
\ead{R.N.Lee@inp.nsk.su}
\fntext[DFG]{Supported by DFG (under the grant No. GZ 436 RUS 113/769/0-2)}
\address[BINP]{Budker Institute of Nuclear Physics, 630090, Novosibirsk, Russia}
\address[NSU]{Novosibirsk State University, 630090, Novosibirsk, Russia}
\date{\today}
\begin{abstract}
We derive the low-energy expansion of $\left(  Z\alpha\right)  ^{2}$ and
$\left(  Z\alpha\right)  ^{4}$ terms of the polarization operator in the
Coulomb field. Physical applications such as the low-energy Delbr\"{u}ck
scattering and "magnetic loop" contribution to the $g$ factor of the bound
electron are considered.
\end{abstract}
\end{frontmatter}

\section{Introduction}

One of the predictions of the quantum field theory is a vacuum polarization by
an external field. An important case thoroughly
studied both experimentally and theoretically is the vacuum polarization
effects in atomic field. Methods used for the study of this effect essentially
depend on the nuclear charge $Z\left\vert e\right\vert $. At low $Z$, the
perturbation theory with respect to $Z\alpha$ is applicable ($\alpha
=e^{2}=1/137$ is the fine structure constant, $\hbar=c=1$). At high $Z$, the
interaction with the external field should be taken into account exactly,
which can be done with the help of the electron Green function in this field.
This approach often requires quite involved numerical calculations, which
usually fail to give the results for low $Z$. Thus, the two approaches tend to
be complementary. Usually, the perturbative calculations of vacuum polarization effect are limited by the
leading order, since the first nonvanishing correction involves two more loops. Nowadays, the modern methods of calculation of the multiloop integrals
are sufficiently powerful for the calculation of higher orders in $Z\alpha$.
It provides a possibility to compare the results of these approaches.

One of the basic nonlinear QED processes in the atomic field is the
Delbr\"{u}ck scattering \cite{Delbruck1933}, the scattering of the photon in
the Coulomb field due to the vacuum polarization. The amplitude of this
process in the Born approximation has been obtained long ago for arbitrary
energies in Ref.\,\cite{Costantini:1971cj}. At high energies and small
scattering angles, when the quasiclassical approximation is valid, the
amplitude is known exactly in $Z\alpha$, see Refs.
\cite{Cheng1969a,Cheng1970,Cheng1972,Milstein1983,Milstein1983a,Lee1999}.
Recently in Ref.\,\cite{Kirilin2008}, the Delbr\"{u}ck amplitude has been
calculated numerically exactly in the parameter $Z\alpha$ at low energies. It
was shown that the contribution of the higher orders (Coulomb corrections) to
the amplitude can be well fitted by the polynomial $C_{4}(Z\alpha)^{4}%
+C_{6}(Z\alpha)^{6}$. The calculation of $(Z\alpha)^{4}$ term in perturbation
theory would provide the independent check of the result of
Ref.\,\cite{Kirilin2008}.

In present paper, we consider the polarization operator $\Pi^{\mu\nu}\left(
\omega,\mathbf{k},\mathbf{q}\right)  $ in the Coulomb field for small external
momenta%
\begin{equation}
\omega\sim\left\vert \mathbf{k}\right\vert \sim\left\vert \mathbf{q}%
\right\vert \sim\lambda m,
\end{equation}
where $\lambda$ is a dimensionless small parameter,%
\begin{equation}
\lambda\ll1\,. \label{eq:lambda}%
\end{equation}
We calculate the expansion of the polarization operator in the Coulomb field
in $\lambda$ and $Z\alpha$ up to the order $\lambda^{4}\left(  Z\alpha\right)
^{4}$. The low-energy Delbr\"{u}ck scattering amplitude is readily expressed
in terms of this operator. This polarization operator is also an essential
ingredient of calculations of different physical observables in atoms, like
Lamb shift and magnetic moment of the bound particle.

The polarization operator in the external Coulomb field is determined as
follows:
\begin{equation}
\Pi^{\mu\nu}\left(  \omega,\mathbf{k},\mathbf{q}\right)  =4\pi ie^{2}\int
d\mathbf{x}d\mathbf{y\,}dt e^{-i\omega t+i\mathbf{kx}-i\mathbf{q y}%
}\left\langle \mathrm{vac}\right\vert \mathrm{T}J^{\mu}\left(  t,\mathbf{x}%
\right)  J^{\nu}\left(  0,\mathbf{y}\right)  \left\vert \mathrm{vac}%
\right\rangle \,,\label{eq:PmunuDefinition}%
\end{equation}
where $J^{\mu}=\bar{\psi}\gamma^\mu\psi$ is the electron current and the state $\left\vert \mathrm{vac}\right\rangle $ corresponds to the vacuum state in the presence of the Coulomb potential $Z\left\vert e\right\vert /r$. In $e^{2}$ order, we have
\begin{equation}
\Pi^{\mu\nu}\left(  \omega,\mathbf{k},\mathbf{q}\right)  =4\pi ie^{2}\int
\frac{d\varepsilon}{2\pi}d\mathbf{x}d\mathbf{y\,} e^{i\mathbf{kx}%
-i\mathbf{qy}}\mathrm{Tr}\left[  \gamma^{\mu}G\left(  \mathbf{x}%
,\mathbf{y}|\varepsilon-\omega\right)  \gamma^{\nu}G\left(  \mathbf{y,x}%
|\varepsilon\right)  \right]  \,,\label{eq:PiviaG}%
\end{equation}
where $G\left(  \mathbf{y,x}|\varepsilon\right)  $ is the Green function of
the electron in the Coulomb field. Due to the gauge invariance, the
polarization operator obeys the constraints%
\begin{equation}
k_{\mu}\Pi^{\mu\nu}\left(  \omega,\mathbf{k},\mathbf{q}\right)  =q_{\nu}%
\Pi^{\mu\nu}\left(  \omega,\mathbf{k},\mathbf{q}\right)  =0,\label{e1:10}%
\end{equation}
where $k^{0}=q^{0}=\omega$. Using these constraints, we can express $\Pi
^{\mu\nu}$ via five independent tensor structures, which we choose as follows
\begin{align}
m^{3}\Pi^{\mu\nu} &  =f_{1}\left(  g^{\mu\nu}\,k\cdot q-q^{\mu}k^{\nu}\right)
-f_{3}\epsilon^{\mu\alpha\beta\gamma}n_{\alpha}\epsilon^{\nu\rho\sigma\tau
}n_{\rho}\,\frac{k_{\beta}\,\left(  k-q\right)  _{\gamma}\left(  k-q\right)
_{\sigma}q_{\tau}}{\left(  k-q\right)  ^{2}}\nonumber\\
&  +\left(  n^{\mu}k^{\alpha}-\omega g^{\mu\alpha}\right)  \left(  q^{\beta
}n^{\nu}-\omega g^{\beta\nu}\right)  \left[  f_{2}g_{\alpha\beta}+f_{4}%
\,\frac{k_{\alpha}q_{\beta}}{\omega^{2}}-f_{5}\frac{\left(  k-q\right)
_{\alpha}\left(  k-q\right)_{\beta}}{(k-q)^{2}}\right]
,\label{eq:Parametrization}%
\end{align}
where $n=\left(  1,\mathbf{0}\right)  $ and $f_{i}$ are some scalar functions
of $\omega$, $\mathbf{k}$ and $\mathbf{q}$.

It is important to note that the low-energy expansion of $\Pi^{\mu\nu}$ itself
and the functions $f_{1-5}$ is not reduced to the multiple Taylor expansion in
$k^{i},q^{i}$ and $\omega$, i.e., $f_{1-5}$ are nonanalytic functions of the
external momenta. Nevertheless, the expansion in powers of the parameter
$\lambda$ from Eq.\,(\ref{eq:lambda}) is still possible. Different terms of
the expansion come from one of two different regions of integration. We
separate the contributions of these regions, using the dimensional
regularization, in spirit of Refs. \cite{Beneke:1997zp,Smirnov:1998vk}. The
methods of separation and calculation of the contributions of these two
regions are given in Sections \ref{Sec:Separation} and \ref{Sec:Calculation},
respectively. The results and conclusions are presented in Section
\ref{Sec:Results}.

\section{Separation of the contributions of soft and hard
regions\label{Sec:Separation}}

\begin{figure}
[ptb]
\begin{center}
\includegraphics[
height=0.8743in,
width=1.8715in
]%
{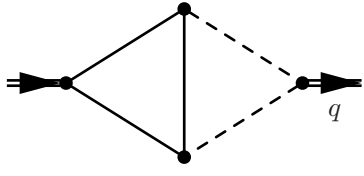}%
\caption{Example of the integral having nonanalytic expansion in $q$.}%
\label{Fig:Example}%
\end{center}
\end{figure}

In order to demonstrate our method, let us consider the behaviour of the
integral
\begin{multline}
J^{\left(  \mathcal{D}\right)  }\left(  \mathbf{q}^{2}\right)  =\int
\frac{d^{\mathcal{D}}\Delta d^{\mathcal{D}}p}{\pi^{\mathcal{D}}}\,j\left(
\mathbf{p},\boldsymbol{\Delta},\mathbf{q}\right) \\
=\int\frac{\pi
^{-\mathcal{D}}d^{\mathcal{D}}\Delta d^{\mathcal{D}}p}{\mathbf{\Delta}%
^{2}\left(  \mathbf{q}+\boldsymbol{\boldsymbol{\Delta}}\right)  ^{2}\left(
\mathbf{p}^{2}+1\right)  \left[  \left(  \mathbf{p}+\mathbf{q}\right)
^{2}+1\right]  \left[  \left(  \mathbf{p}-\boldsymbol{\boldsymbol{\Delta}%
}\right)  ^{2}+1\right]  },
\end{multline}
at $\mathbf{q}^{2}\ll1$. The corresponding diagram is depicted in
Fig.\thinspace\ref{Fig:Example}.
The small-$\mathbf{q}^{2}$ expansion of this integral, obtained in
Ref.\thinspace\cite{Broadhurst:1993mw}, has the form:%
\begin{equation}
J^{\left(  \mathcal{D}\right)  }\left(  \mathbf{q}^{2}\right)  =\sum
_{n=0}^{\infty}C_{n}\left(  \mathcal{D}\right)  \,\left(  -\mathbf{q}%
^{2}\right)  ^{n}+\left(  \mathbf{q}^{2}\right)  ^{\frac{\mathcal{D}}{2}%
-2}\sum_{n=0}^{\infty}D_{n}\left(  \mathcal{D}\right)  \,\left(
-\mathbf{q}^{2}\right)  ^{n}\,, \label{e2:3}%
\end{equation}%
\begin{align}
C_{n}\left(  \mathcal{D}\right)   &  =\frac{\Gamma\left(  n+3-\mathcal{D}%
/2\right)  }{\left(  n+3-\mathcal{D}/2\right)  \left(  \mathcal{D}-3\right)
}\left[  \Gamma\left(  \mathcal{D}/2-1\right)  \frac{\Gamma\left(
n+5-\mathcal{D}\right)  \Gamma\left(  n+3-\mathcal{D}/2\right)  }%
{\Gamma\left(  n+\mathcal{D}/2\right)  \Gamma\left(  2n+6-\mathcal{D}\right)
}\right. \nonumber\\
&  \left.  -\frac{\Gamma\left(  n+2\right)  \Gamma\left(  2-\mathcal{D}%
/2\right)  }{\Gamma\left(  2n+3\right)  }\right]  ,
\end{align}%
\begin{equation}
D_{n}\left(  \mathcal{D}\right)  =\Gamma\left(  2-\mathcal{D}/2\right)
\Gamma\left(  \mathcal{D}/2-1\right)  ^{2}\frac{\Gamma\left(  n+1\right)
\Gamma\left(  n+3-\mathcal{D}/2\right)  }{\Gamma(\mathcal{D}-2)\Gamma\left(
2n+3\right)  }\,.
\end{equation}
The representation (\ref{e2:3}) determines $J^{\left(  \mathcal{D}\right)
}\left(  \mathbf{q}^{2}\right)  $ at $0<\mathbf{q}^{2}<4.$ The limit
$\mathbf{q}^{2}\rightarrow0$ essentially depends on $\mathcal{D}$. For
$\mathcal{D}>4$, this limit is equal to $C_{0}\left(  \mathcal{D}\right)  $
and can be considered as the value of the function $J^{\left(  \mathcal{D}%
\right)  }$ at $\mathbf{q}^{2}=0.$ For $\mathcal{D}<4$, there is no finite
$\mathbf{q}^{2}\rightarrow0$ limit of the representation (\ref{e2:3}).
However, the value $J^{\left(  \mathcal{D}\right)  }\left(  0\right)  $
defined via the analytic continuation with respect to $\mathcal{D}$ from the
region $\mathcal{D}>4$ is still $C_{0}\left(  \mathcal{D}\right)  .$ In other
words, the limit $\mathbf{q}^{2}\rightarrow0$ is not commuting with the
analytic continuation with respect to $\mathcal{D}$. The same claim is valid
for the derivatives of $J^{\left(  \mathcal{D}\right)  }\left(  \mathbf{q}%
^{2}\right)  $ with respect to $\mathbf{q}^{2}$. As a consequence, the
integrand expanded in $\mathbf{q}$ gives only the terms $\propto\left(
\mathbf{q}^{2}\right)  ^{n}$ after the integration within the dimensional
regularization. These terms correspond to the first sum in the right-hand side
of Eq.\thinspace(\ref{e2:3}):
\begin{equation}
\int\frac{d^{\mathcal{D}}\Delta\,d^{\mathcal{D}}p}{\pi^{\mathcal{D}}}\left[
\sum_{n=0}^{\infty}\frac{q^{i_{1}}\ldots q^{i_{n}}}{n!}\left.  \frac
{\partial^{n}\,j\left(  \mathbf{p},\boldsymbol{\Delta},\mathbf{q}\right)
}{\partial q^{i_{1}}\ldots\partial q^{i_{n}}}\right\vert _{\mathbf{q}%
=0}\right]  =\sum_{n=0}^{\infty}C_{n}\left(  \mathcal{D}\right)  \,\left(
-\frac{\mathbf{q}^{2}}{4}\right)  ^{n}.
\end{equation}
Note that the expansion of the massless propagators is valid only in the
region $\Delta\gg q$ (hard region). In order to obtain the rest terms of the
expansion (\ref{e2:3}) one has to determine the contribution of the region
$\Delta\sim q$ (soft region). To separate these contributions, we use the
following trick. In the soft\emph{ }region the massive propagators \ can be
expanded in both $\mathbf{q}$ and $\boldsymbol{\Delta}$. Let us truncate this
expansion at some fixed order $N$ in $\mathbf{q}$ and $\boldsymbol{\Delta}$:
\begin{align}
j_{\mathrm{soft}}\left(  N,\mathbf{p},\boldsymbol{\Delta},\mathbf{q}\right)
&  =\sum_{n=0}^{N}j_{\mathrm{soft}}^{(n)}\left(  \mathbf{p},\boldsymbol{\Delta
},\mathbf{q}\right)  ,\\
j_{\mathrm{soft}}^{(n)}\left(  \mathbf{p},\boldsymbol{\Delta},\mathbf{q}%
\right)   &  =\left.  \frac{\partial^{n}\tau^{4}j\left(  \mathbf{p}%
,\tau\boldsymbol{\Delta},\tau\mathbf{q}\right)  }{n!\partial\tau^{n}%
}\right\vert _{\tau=0}\,,\text{ so that }j_{\mathrm{soft}}^{(n)}=O\left(
q^{n-4}\right)  .\nonumber
\end{align}
Now we can identically rewrite $J^{\left(  \mathcal{D}\right)  }\left(
\mathbf{q}^{2}\right)  $ as
\begin{align}
& J^{\left(  \mathcal{D}\right)  }\left(  \mathbf{q}^{2}\right)
=J_{\mathrm{hard}}^{\left(  \mathcal{D}\right)  }\left(  \mathbf{q}%
^{2}\right)  +J_{\mathrm{soft}}^{\left(  \mathcal{D}\right)  }\left(
\mathbf{q}^{2}\right) \nonumber\\
&  =\int\frac{d^{\mathcal{D}}\Delta\,d^{\mathcal{D}}p}{\pi^{\mathcal{D}}%
}\,\left[  j\left(  \mathbf{p},\boldsymbol{\Delta},\mathbf{q}\right)
-\,j_{\mathrm{soft}}\left(  N,\mathbf{p},\boldsymbol{\Delta},\mathbf{q}%
\right)  \right]  +\int\frac{d^{\mathcal{D}}\Delta\,d^{\mathcal{D}}p}%
{\pi^{\mathcal{D}}}\,j_{\mathrm{soft}}\left(  N,\mathbf{p},\boldsymbol{\Delta
},\mathbf{q}\right)  . \label{eq:add and subtract}%
\end{align}
The contribution of the soft region in the first term in Eq.\thinspace
(\ref{eq:add and subtract}) is suppressed as $q^{\mathcal{D+}N-3}$, thus the
integral of the difference $j\left(  \mathbf{p},\boldsymbol{\Delta}%
,\mathbf{q}\right)  -\,j_{\mathrm{soft}}\left(  N,\mathbf{p}%
,\boldsymbol{\Delta},\mathbf{q}\right)  $ is determined by the region
$\Delta\sim1\gg q$ up to $O\left(  q^{2\left\lfloor \left(  \mathcal{D+}%
N-3\right)  /2\right\rfloor }\right)  $ term. In fact, the expansion of
$\,j_{\mathrm{soft}}\left(  N,\mathbf{p},\boldsymbol{\Delta},\mathbf{q}%
\right)  $ in $\mathbf{q}$ gives scaleless functions of $\boldsymbol{\Delta}$,
which vanish after the integration in the dimensional regularization. Finally,
we have
\begin{align}
J^{\left(  \mathcal{D}\right)  }\left(  \mathbf{q}^{2}\right)   &
=J_{\mathrm{hard}}^{\left(  \mathcal{D}\right)  }\left(  \mathbf{q}%
^{2}\right)  +J_{\mathrm{soft}}^{\left(  \mathcal{D}\right)  }\left(
\mathbf{q}^{2}\right) \nonumber\\
&  =\sum_{n=0}^{\infty}\int\frac{d^{\mathcal{D}}\Delta\,d^{\mathcal{D}}p}%
{\pi^{\mathcal{D}}}\,j_{\mathrm{hard}}^{(n)}\left(  \mathbf{p}%
,\boldsymbol{\Delta},\mathbf{q}\right)  +\sum_{n=0}^{\infty}\int
\frac{d^{\mathcal{D}}\Delta\,d^{\mathcal{D}}p}{\pi^{\mathcal{D}}%
}\,j_{\mathrm{soft}}^{(n)}\left(  \mathbf{p},\boldsymbol{\Delta}%
,\mathbf{q}\right)  ,\\
j_{\mathrm{hard}}^{(n)}\left(  \mathbf{p},\boldsymbol{\Delta},\mathbf{q}%
\right)   &  =\left.  \frac{\partial^{n}j\left(  \mathbf{p},\boldsymbol{\Delta
},\tau\mathbf{q}\right)  }{n!\partial\tau^{n}}\right\vert _{\tau
=0},~j_{\mathrm{soft}}^{(n)}\left(  \mathbf{p},\boldsymbol{\Delta}%
,\mathbf{q}\right)  =\left.  \frac{\partial^{n}\tau^{4}j\left(  \mathbf{p}%
,\tau\boldsymbol{\Delta},\tau\mathbf{q}\right)  }{n!\partial\tau^{n}%
}\right\vert _{\tau=0}.\nonumber
\end{align}

Let us now use this approach for the calculation of the low-energy expansion
of the polarization operator in the Coulomb field.%
\begin{figure}
[ptb]
\begin{center}
\includegraphics[
height=1.7002in,
width=1.7737in
]%
{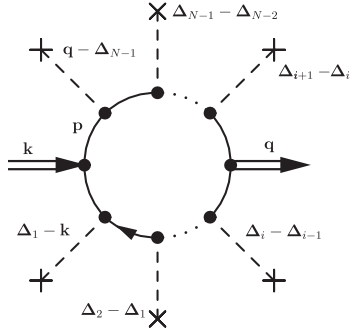}%
\caption{The $(Z\alpha)^{N}$ contribution to the polarization operator. Solid
lines denote the electron propagator, dashed lines denote the Coulomb field.}%
\label{fig:DelbruckN}%
\end{center}
\end{figure}
The $(Z\alpha)^{N}$ contribution ($N$ is even) is determined by the $N$-loop
diagrams depicted in Fig.\thinspace\thinspace\ref{fig:DelbruckN}. In the
dimensional regularization, it can be represented in the form%
\begin{equation}
\Pi_{(Z\alpha)^{N}}^{\mu\nu}\left(  \omega,\mathbf{k},\mathbf{q}\right)
=\int\frac{d\varepsilon\,d^{\mathcal{D}}p}{\left(  2\pi\right)  ^{\mathcal{D}%
+1}}\frac{\prod_{i=1}^{N-1}d^{\mathcal{D}}\Delta_{i}}{\left(  2\pi\right)
^{\left(  N-1\right)  \mathcal{D}}}\,\mathfrak{P}^{\mu\nu}\left(
\varepsilon,\mathbf{p},\boldsymbol{\Delta}_{1},\ldots\boldsymbol{\Delta}%
_{N-1},\omega,\mathbf{k},\mathbf{q}\right)  . \label{eq:PolarizationOperatorN}%
\end{equation}
Similar to the previous example, there are two different region of integration%
\begin{align}
\text{\textit{hard} region, when }  &  \varepsilon\sim p^{i}\sim
\Delta_{1..N-1}^{i}\sim m,\label{e1:2}\\
\text{\textit{soft} region, when }  &  \left\{
\begin{array}
[c]{rl}%
\varepsilon\sim p^{i} & \sim m,\\
\Delta_{1..N-1}^{i} & \sim\lambda\,m.
\end{array}
\right.  \label{e1:3}%
\end{align}
Again, the expansion of the polarization operator is the sum of the integrals
of the expansion of the integrand in hard and soft regions. In the coordinate
representation, these regions have a simple physical meaning. The
characteristic size of the electron field fluctuations (the size of the
electron loop) is of the order $1/m$. Hard region corresponds to the
configurations where the distance between the Coulomb source and the electron
loop is of the order of $1/m$. Soft region corresponds to the creation of the
virtual electron-positron pair far from the Coulomb source. Obviously, there
is no contribution from the region where only some of the momenta
$\boldsymbol{\Delta}_{n}$ are hard while the rest are soft. In the momentum
representation, these regions correspond to massless tadpole diagrams which
are zero in the dimensional regularization. Thus, the expansion of the
polarization operator (\ref{eq:PolarizationOperatorN}) has the form%
\begin{multline}
\Pi_{(Z\alpha)^{N}}^{\mu\nu}\left(  \omega,\mathbf{k},\mathbf{q}\right)
=\Pi_{(Z\alpha)^{N},\text{hard}}^{\mu\nu}\left(  \omega,\mathbf{k}%
,\mathbf{q}\right)  +\Pi_{(Z\alpha)^{N},\text{soft}}^{\mu\nu}\left(
\omega,\mathbf{k},\mathbf{q}\right) \\
 =\sum_{n}\int\frac{d\varepsilon\,d^{\mathcal{D}}p}{\left(  2\pi\right)
^{\mathcal{D}+1}}\frac{\prod_{i=1}^{N-1}d^{\mathcal{D}}\Delta_{i}}{\left(
2\pi\right)  ^{\left(  N-1\right)  \mathcal{D}}}\,\mathfrak{P}_{\text{hard}%
}^{\mu\nu\left(  n\right)  }+\sum_{n}\int\frac{d\varepsilon\,d^{\mathcal{D}}%
p}{\left(  2\pi\right)  ^{\mathcal{D}+1}}\frac{\prod_{i=1}^{N-1}%
d^{\mathcal{D}}\Delta_{i}}{\left(  2\pi\right)  ^{\left(  N-1\right)
\mathcal{D}}}\,\mathfrak{P}_{\text{soft}}^{\mu\nu\left(  n\right)
},\label{eq:POHardSoft}
\end{multline}
\begin{align}
\mathfrak{P}_{\text{hard}}^{\mu\nu\left(  n\right)  }  &  =\left.
\frac{\partial^{n}\mathfrak{P}^{\mu\nu}\left(  \varepsilon,\mathbf{p}%
,\boldsymbol{\Delta}_{1},\ldots\boldsymbol{\Delta}_{N-1},\tau\omega
,\tau\mathbf{k},\tau\mathbf{q}\right)  }{n!\partial\tau^{n}}\right\vert
_{\tau=0},\\
\mathfrak{P}_{\text{soft}}^{\mu\nu\left(  n\right)  }  &  =\left.
\frac{\partial^{n}\tau^{N-2}\mathfrak{P}^{\mu\nu}\left(  \varepsilon
,\mathbf{p},\tau\boldsymbol{\Delta}_{1},\ldots\tau\boldsymbol{\Delta}%
_{N-1},\tau\omega,\tau\mathbf{k},\tau\mathbf{q}\right)  }{n!\partial\tau^{n}%
}\right\vert _{\tau=0}.
\end{align}
Note that the simple power counting allows one to estimate the leading terms
of the hard and soft contribution as%
\begin{align}
\Pi_{(Z\alpha)^{N},\text{hard}}^{\mu\nu}\left(  \omega,\mathbf{k}%
,\mathbf{q}\right)   &  \sim\lambda^{2},\nonumber\\
\Pi_{(Z\alpha)^{N},\text{soft}}^{\mu\nu}\left(  \omega,\mathbf{k}%
,\mathbf{q}\right)   &  \sim\lambda^{\left(  N-1\right)  \left(
\mathcal{D}-1\right)  +1}.
\end{align}
Using Eq.\thinspace(\ref{eq:POHardSoft}), one can calculate the contributions
of the hard and soft regions separately.

\section{Method of calculation\label{Sec:Calculation}}%

\begin{figure}
[ptb]
\begin{center}
\includegraphics[
height=1.0438in,
width=1.1467in
]%
{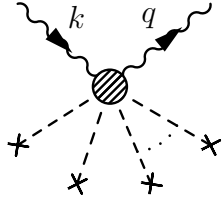}%
\caption{Graphical representation of the soft contribution.}%
\label{fig:HELVertex}%
\end{center}
\end{figure}
The contribution of the soft region can be graphically represented as the tree
diagram shown in Fig.\thinspace\ref{fig:HELVertex}. The local multiphoton
vertex depicted as a thick dot corresponds to the expansion of the fermionic
loop with respect to the soft momenta $\boldsymbol{\Delta}_{i},k,q$. The
expansion is expressed via the integrals of the following form%
\begin{equation}
\int\frac{d\varepsilon\,d^{\mathcal{D}}p}{\left(  2\pi\right)  ^{\mathcal{D}%
+1}}\left(  \varepsilon^{2}-\mathbf{p}^{2}-m^{2}+i0\right)  ^{-n}%
=\frac{i\left(  -1\right)  ^{n}\Gamma\left(  n-\mathcal{D}/2-1/2\right)
}{\left(  4\pi\right)  ^{\left(  \mathcal{D}+1\right)  /2}\Gamma\left(
n\right)  m^{2n-\mathcal{D}-1}}\,.
\end{equation}
The remaining integrals over $\boldsymbol{\Delta}_{i}$ can be easily evaluated
in the coordinate representation. Naturally, the contribution of the soft
region can be calculated also with the help of the derivative expansion of the
one-loop effective QED action.

The contribution of the hard region determined by Eq.\thinspace
(\ref{eq:POHardSoft}) is expressed in terms of the $N$-loop tadpoles. In
particular, in $\left(  Z\alpha\right)  ^{2}$ order, the basic integral has
the following form%
\begin{equation}
I=\int\frac{d\varepsilon\,d^{\mathcal{D}}p\,d^{\mathcal{D}}\Delta}{\left(
2\pi\right) ^{2\mathcal{D}+1}}\frac{1}{\left(  \varepsilon^{2}-\mathbf{p}^{2}%
-m^{2}+i0\right)^{n_{1}}\left(  \varepsilon^{2}-\left(  \mathbf{p-}%
\boldsymbol{\Delta}\right)^{2}-m^{2}+i0\right)^{n_{2}}\left(
\boldsymbol{\Delta}^{2}\right)^{n_{3}}}.
\end{equation}
After the Wick rotation $\varepsilon\rightarrow i\varepsilon$ and rescaling
$\mathbf{p}\rightarrow\sqrt{\varepsilon^{2}+m^{2}}\mathbf{p},\mathbf{~}%
\boldsymbol{\Delta}\rightarrow\sqrt{\varepsilon^{2}+m^{2}}\boldsymbol{\Delta}%
$, we integrate over $\varepsilon$ and obtain%
\begin{align}
I  &  =i\frac{m^{1-2\gamma}\Gamma\left(
\gamma-1/2\right)  }{2\sqrt{\pi}\,\Gamma\left(  \gamma\right)  }\int
\frac{d^{\mathcal{D}}p\,d^{\mathcal{D}}\Delta}{\left(  2\pi\right)
^{2\mathcal{D}}}\frac{\left(  -1\right)  ^{n_{1}+n_{2}}}{\left(\mathbf{p}^{2}+1\right)^{n_{1}}\left(  \left(
\mathbf{p-}\boldsymbol{\Delta}\right)  ^{2}+1\right)^{n_{2}}\left(
\boldsymbol{\Delta}^{2}\right)^{n_{3}}},\\
\gamma &  =n_{1}+n_{2}+n_{3}-\mathcal{D}.\nonumber
\end{align}
The remaining integral is the two-loop tadpole in $\mathcal{D}=3+\epsilon$
dimensions which can be easily expressed in terms of $\Gamma$-functions.

\begin{figure}
[ptb]
\begin{center}
\includegraphics[
trim=0.000000in 0.000000in -0.034993in 0.000000in,
height=1.0032in,
width=1.0032in
]%
{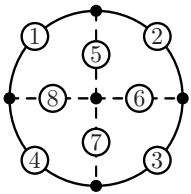}%
\caption{Topology of the integrals required for the calculation of the
low-energy expansion of the polarization operator in $\left(  Z\alpha\right)
^{4}$ order.}%
\label{Fig:Topology}%
\end{center}
\end{figure}

Performing the similar integration over $\varepsilon$ in $\left(
Z\alpha\right)  ^{4}$ order, we express the contribution of the hard region in
terms of the integrals of the topology (and its subtopologies) depicted in
Fig.\,\ref{Fig:Topology}.
The general form of such integral is%
\begin{equation}
J_{n_{1}\ldots n_{10}}=\int\frac{d^{\mathcal{D}}k_{1}d^{\mathcal{D}}%
k_{2}d^{\mathcal{D}}k_{3}d^{\mathcal{D}}k_{4}D_{9}^{-n_{9}}D_{10}^{-n_{10}}%
}{\pi^{2\mathcal{D}}D_{1}^{n_{1}}D_{2}^{n_{2}}D_{3}^{n_{3}}D_{4}^{n_{4}}%
D_{5}^{n_{5}}D_{6}^{n_{6}}D_{7}^{n_{7}}D_{8}^{n_{8}}},
\end{equation}
where%
\begin{gather}
D_{1,\ldots,4}=k_{1,\ldots,4}^{2}+1,\qquad\ \\
D_{5}=\left(  k_{1}-k_{2}\right)  ^{2},\quad D_{6}=\left(  k_{2}-k_{3}\right)
^{2},\nonumber\\
D_{7}=\left(  k_{3}-k_{4}\right)  ^{2},\quad D_{8}=\left(  k_{4}-k_{1}\right)
^{2},\nonumber\\
D_{9}=\left(  k_{1}-k_{3}\right)  ^{2},\quad D_{10}=\left(  k_{2}%
-k_{4}\right)  ^{2}.\nonumber
\end{gather}

\begin{figure}[ptb]
\subfigure[$j_{\text{Cake}}$]{\includegraphics[clip,width=2cm]{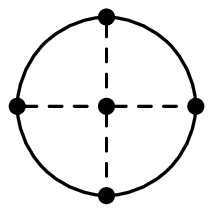}}\hfill
{}%
\subfigure[$j_{\text{Infinity}}$]{\includegraphics[width=2cm]{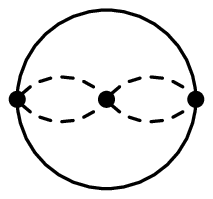}}\hfill
{}\subfigure[$j_{\text{Melon}}$]{\includegraphics[width=2cm]{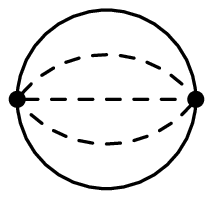}}\hfill
{}%
\subfigure[$j_{\text{Tumbler}}$]{\includegraphics[width=2cm]{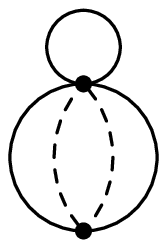}}\hfill
{}%
\subfigure[$j_{\text{Clover}}$]{\includegraphics[width=2cm]{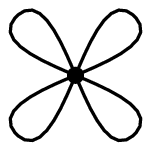}}\caption{Master
integrals of the topology, shown in Fig.\,\ref{Fig:Topology}.}%
\label{fig:MIs}%
\end{figure}

The IBP reduction procedure {\cite{Chetyrkin1980,Chetyrkin1981}} allows one to
express any vacuum integral of the considered topology via the five master
integrals shown in Fig.\,\ref{fig:MIs}. On these diagrams the solid and dashed lines denote the massive
$\left(  k^{2}+1\right)  ^{-1}$ and massless $\left(  k^{2}\right)  ^{-1}$
propagators. For each loop momentum the integration measure is taken as
\[
\frac{d^{\mathcal{D}}k}{\pi^{\mathcal{D}/2}}.
\]
Four of these integrals are trivially expressed in terms of $\Gamma
$-functions. Their explicit forms are presented in Appendix. The only
nontrivial master integral is $J_{\text{Cake}}$. After the IBP reduction, the
master integral $J_{\text{Cake}}$ enters the polarization operator with the
coefficient, having the first-order pole in the point $\mathcal{D}=3$.
Therefore, we need to determine the $O\left(  \epsilon^{0}\right)  $ and
$O\left(  \epsilon^{1}\right)  $ terms of the expansion of $J_{\text{Cake}}$.

We find it convenient to use the recurrence relation with respect to
space-time dimension, see Ref.\thinspace{\cite{Tarasov1996a}}. First, we use
the Feynman parameterization to obtain the relation%
\begin{align}
J_{\text{Cake}}^{(\mathcal{D}-2)}\overset{def}{\equiv}J_{1111111100}%
^{(\mathcal{D}-2)}  &  =8J_{1112122200}^{(\mathcal{D})}+8J_{1112221200}%
^{(\mathcal{D})}+8J_{1122211200}^{(\mathcal{D})}+8J_{1222111200}%
^{(\mathcal{D})}\nonumber\\
&  +4J_{1122121200}^{(\mathcal{D})}+4J_{1212112200}^{(\mathcal{D}%
)}+4J_{1212121200}^{(\mathcal{D})}+J_{2222111100}^{(\mathcal{D})}.
\label{eq:recurrence}%
\end{align}
Then, using the IBP identities, we express the integrals in the right-hand
side of this relation via the five master integrals from Fig.\thinspace
\ref{fig:MIs}. In particular, the last term in Eq.\thinspace
(\ref{eq:recurrence}) can be expressed via the master integrals as follows%
\begin{align}
J_{2}^{(5+\epsilon)}\overset{def}{\equiv}J_{2222111100}^{(5+\epsilon)}  &
=a_{\text{Cake}}J_{\text{Cake}}^{(5+\epsilon)}+\frac{a_{\text{Clover}}%
}{\epsilon^{3}}J_{\text{Clover}}^{(5+\epsilon)}\nonumber\\
&  +a_{\text{Infinity}}J_{\text{Infinity}}^{(5+\epsilon)}+\frac
{a_{\text{Tumbler}}}{\epsilon^{2}}J_{\text{Tumbler}}^{(5+\epsilon)}%
+\frac{a_{\text{Melon}}}{\epsilon^{2}}J_{\text{Melon}}^{(5+\epsilon)}\,.
\label{eq:j2222}%
\end{align}
The coefficients $a_{i}$ are presented in the Appendix. They are chosen to be
finite in the limit $\epsilon\rightarrow0$. After the reduction, we have the
following recurrence relation:%
\begin{multline}
J_{\text{Cake}}^{(3+\epsilon)}=\epsilon b_{\text{Cake}}J_{\text{Cake}%
}^{(5+\epsilon)}+\frac{b_{\text{Clover}}}{\epsilon^{2}}\,J_{\text{Clover}%
}^{(5+\epsilon)}\\+b_{\text{Infinity}}J_{\text{Infinity}}^{(5+\epsilon)}%
+\frac{b_{\text{Tumbler}}}{\epsilon}\,J_{\text{Tumbler}}^{(5+\epsilon)}%
+\frac{b_{\text{Melon}}}{\epsilon}\,J_{\text{Melon}}^{(5+\epsilon)}\,.
\end{multline}
Again, the coefficients $b_{i}$ are chosen to be finite in the limit
$\epsilon\rightarrow0$ and are presented in the Appendix. Now we use the
following trick. Let us express $J_{\text{Cake}}^{(5+\epsilon)}$ from
Eq.\thinspace(\ref{eq:j2222}) and substitute into Eq.\thinspace
\ (\ref{eq:recurrence}). We obtain%
\begin{multline}
J_{\text{Cake}}^{(3+\epsilon)}  =\frac{\epsilon b_{\text{Cake}}%
}{a_{\text{Cake}}}\,J_{2}^{(5+\epsilon)}+\frac{b_{\text{Cake}}}{\epsilon^{2}%
}\left(  \frac{b_{\text{Clover}}}{b_{\text{Cake}}}-\frac{a_{\text{Clover}}%
}{a_{\text{Cake}}}\right)  J_{\text{Clover}}^{(5+\epsilon)}\\
+b_{\text{Cake}%
}\left(  \frac{b_{\text{Infinity}}}{b_{\text{Cake}}}-\frac{\epsilon
\,a_{\text{Infinity}}}{a_{\text{Cake}}}\right)  J_{\text{Infinity}%
}^{(5+\epsilon)}+\frac{b_{\text{Cake}}}{\epsilon}\left(  \frac{b_{\text{Tumbler}}%
}{b_{\text{Cake}}}\,-\frac{a_{\text{Tumbler}}}{a_{\text{Cake}}}\right)
J_{\text{Tumbler}}^{(5+\epsilon)}\\
+\frac{b_{\text{Cake}}}{\epsilon}\left(
\frac{b_{\text{Melon}}}{b_{\text{Cake}}}-\frac{a_{\text{Melon}}}%
{a_{\text{Cake}}}\right)  J_{\text{Melon}}^{(5+\epsilon)}\,.
\label{eq:jCakeviaj2D}%
\end{multline}
Since the integral $J_{2}^{\left(  \mathcal{D}\right)  }$ is finite in
$\mathcal{D}=5$ and the coefficient in front of this integral in
Eq.\thinspace(\ref{eq:jCakeviaj2D}) contains $\epsilon$ factor, the first term
in the right-hand side of Eq.\thinspace(\ref{eq:jCakeviaj2D}) does not
contribute in $\epsilon^{0}$ order. Expanding the coefficient $a_{i},b_{i}$
and the four simple master integrals, we obtain%
\begin{equation}
J_{\text{Cake}}^{(3+\epsilon)}=\frac{\pi^{4}}{6}+\epsilon\left[  \frac{\pi
^{4}}{3}\left(  C-\ln2-\frac{11}{8}\right)  -\pi^{2}-\frac{3}{4}\,J_{2}%
^{(5)}\right]  +O\left(  \epsilon^{2}\right)  \,, \label{eq:jCakeviaj2}%
\end{equation}
where $C=0.577...$ is the Euler constant. Note, that using this trick we have
obtained the $O\left(  \epsilon^{0}\right)  $ term of $J_{\text{Cake}%
}^{(3+\epsilon)}$ {}\textquotedblleft for free\textquotedblright.

In order to calculate the $O\left(  \epsilon\right)  $ term, let us consider
the general solution of the recurrence (\ref{eq:recurrence}). Taking into
account the explicit form of the coefficient $b_{\text{Cake}}$, we obtain%
\begin{equation}
J_{\text{Cake}}^{(\mathcal{D})}=J_{0}\left(  \mathcal{D}\right)  \left[
P\left(  \mathcal{D}\right)  +\sum_{i=1}^{4}\sum_{n=1}^{\infty}\frac
{c_{i}^{(\mathcal{D}+2n)}}{J_{0}(\mathcal{D}+2n)}J_{i}^{(\mathcal{D}%
+2n)}\right]  , \label{eq:solutionDrelation}%
\end{equation}
where%
\begin{align}
J_{0}\left(  \mathcal{D}\right)   &  =\frac{2^{3\mathcal{D}}\Gamma\left(
1-\mathcal{D}/2\right)  \Gamma\left(  3\mathcal{D}/2-11/2\right)  }%
{\Gamma(\mathcal{D}-2)\Gamma(\mathcal{D}-3)^{2}},\\
c_{\text{Clover}}  &  =\frac{b_{\text{Clover}}}{\epsilon^{3}b_{\text{Cake}}%
}\ ,\quad c_{\text{Infinity}}=\frac{b_{\text{Infinity}}}{\epsilon
b_{\text{Cake}}},\nonumber\\
c_{\text{Tumbler}}  &  =\frac{b_{\text{Tumbler}}}{\epsilon^{2}b_{\text{Cake}}%
}\ ,\quad c_{\text{Melon}}=\frac{b_{\text{Melon}}}{\epsilon^{2}b_{\text{Cake}%
}},\nonumber
\end{align}
and $P\left(  \mathcal{D}\right)  =P\left(  \mathcal{D}+2\right)  $ is a
periodic function of $\mathcal{D}$. Note that, using the explicit form of the
simple master integrals, Eq.\thinspace(\ref{eq:MIsimple}), the sums in
Eq.\thinspace(\ref{eq:solutionDrelation}) can be checked to converge rapidly.
In order to fix the function $P\left(  \mathcal{D}\right)  $, we have to
calculate the leading asymptotic of $J_{\text{Cake}}^{(\mathcal{D})}$ at
$\mathcal{D}\rightarrow+\infty$. However, the calculation of this asymptotic
is not a simple problem. Instead, we may proceed in the alternative way by
applying the method of difference equations described in Ref.\thinspace
{\cite{Laporta2000}}. According to this method, we derive the recurrence
relation in $x$ for $J_{x111111100}^{(\mathcal{D})}$:%
\begin{align}
J_{x111111100}^{(\mathcal{D})}  &  =C(x)J_{x+1,111111100}^{(\mathcal{D}%
)}+F(x),\label{eq:Xrecurrencerelation}\\
C(x)  &  =\frac{x\left(  x+11-3\mathcal{D}\right)  }{\left(  x+5-3\mathcal{D}%
/2\right)  \left(  x+3-\mathcal{D}\right)  }\,,
\end{align}
where $F(x)$ is expressed in terms of finite sums of $\Gamma$- functions. The
solution of this relation is%
\begin{align}
J_{x111111100}^{(\mathcal{D})}  &  =G^{(\mathcal{D})}\left(  x\right)  \left[
R^{(\mathcal{D})}\left(  x\right)  +\sum_{y=x}^{\infty}\frac{F(y)}{G\left(
y\right)  }\right]  ,\label{eq:solutionXrelation}\\
G^{(\mathcal{D})}\left(  x\right)   &  =\frac{\Gamma\left(  x+5-3\mathcal{D}%
/2\right)  \Gamma\left(  x+3-\mathcal{D}\right)  }{\Gamma\left(  x\right)
\Gamma\left(  x+11-3\mathcal{D}\right)  },
\end{align}
where $R^{(\mathcal{D})}\left(  x\right)  $ is a periodic function of $x$,
which can be determined through the $J_{x111111100}^{(\mathcal{D})}$
asymptotic behaviour in $x\rightarrow\infty$ limit. At large $x$, the
$x$-dependence of $J_{x111111100}^{(\mathcal{D})}$ factorizes into $\int
d^{\mathcal{D}}k_{1}\,\left(  k_{1}^{2}+1\right)  ^{-x}$. Using the large-$x$
behaviour of the function $G^{(\mathcal{D})}\left(  x\right)  $, we find%
\begin{equation}
\left.  \frac{J_{x111111100}^{(\mathcal{D})}}{G^{(\mathcal{D})}\left(
x\right)  }\right\vert _{x\rightarrow\infty}\sim x^{3-\mathcal{D}}.
\end{equation}
Therefore, for $\mathcal{D}>3$ we have $R\left(  x\right)  =0$ and we can use
Eq.\thinspace(\ref{eq:solutionXrelation}) to estimate numerically the function
$P\left(  \mathcal{D}\right)  $ in Eq.\thinspace(\ref{eq:solutionDrelation}).
It should be noted that the recurrence relation (\ref{eq:Xrecurrencerelation})
can be hardly applied near $\mathcal{D}=3$ due to the slow convergence of the
sum in the right-hand side of Eq.\thinspace(\ref{eq:solutionXrelation}).
Performing the estimation of $P\left(  \mathcal{D}\right)  $ for several
non-integer values of $\mathcal{D}$, we find that in all cases the value of
the function $P\left(  \mathcal{D}\right)  $ is compatible with zero up to
$10^{-10}$. Thus, our ansatz is $P\left(  \mathcal{D}\right)  =0$, and we have%
\begin{equation}
J_{\text{Cake}}^{(3+\epsilon)}=\frac{8^{\epsilon+3}\Gamma\left(
-\frac{\epsilon}{2}-\frac{1}{2}\right)  \Gamma\left(  \frac{3\epsilon}%
{2}-1\right)  }{\epsilon\Gamma(\epsilon)^{3}}\sum_{n=0}^{\infty}T\left(
2n+\epsilon\right)  , \label{eq:JD}%
\end{equation}%
\begin{align}
T\left(  \nu\right)   &  =\frac{8^{-\nu-5}(\nu+1)\left(  14\nu^{3}+40\nu
^{2}+35\nu+10\right)  \Gamma\left(  -\frac{\nu}{2}-\frac{1}{2}\right)
^{3}\Gamma(\nu)^{3}}{\Gamma\left(  \frac{3}{2}\nu+2\right)  }\nonumber\\
&  -\frac{\pi^{7/2}\left(  6417\nu^{5}+13266\nu^{4}+8368\nu^{3}+980\nu
^{2}-585\nu-126\right)  \Gamma(\nu)}{6144\nu(3\nu-1)(3\nu+1)\Gamma
(-2\nu)\Gamma(3\nu+4)\cos\left(  \frac{\pi\nu}{2}\right)  \sin^{2}(\pi\nu
)\cos\left(  \frac{3\pi\nu}{2}\right)  }\nonumber\\
&  +\frac{\pi^{5/2}\left(  585\nu^{3}+372\nu^{2}+7\nu-12\right)  \Gamma
(\nu)\Gamma\left(  \frac{\nu+1}{2}\right)  ^{2}\tan\left(  \frac{3\pi\nu}%
{2}\right)  }{12288(3\nu-1)(3\nu+1)\Gamma(2\nu+2)\cos\left(  \frac{\pi\nu}%
{2}\right)  \sin\left(  2\pi\nu\right)  }\nonumber\\
&  -\frac{\pi^{3/2}\left(  3897\nu^{4}+3870\nu^{3}+797\nu^{2}-198\nu
-54\right)  \Gamma(\nu)^{2}\Gamma\left(  \frac{\nu+1}{2}\right)  ^{3}%
\tan\left(  \frac{3\pi\nu}{2}\right)  }{36864(3\nu-1)(3\nu+1)\Gamma
(2\nu+2)\Gamma\left(  \frac{3}{2}\nu+\frac{3}{2}\right)  \sin\left(  2\pi
\nu\right)  \sin\left(  \frac{\pi\nu}{2}\right)  }\,.
\end{align}
The series in this representation are converging rapidly. The first term in
$T\left(  \nu\right)  $, the most slowly decreasing one, behaves as $3^{-3n}$.
Thus, roughly speaking, each two consecutive terms in the sum give three more
decimal digits of precision. We claim this expression to be valid for
arbitrary $\epsilon$. For example, one can easily reproduce terms of the
expansion of $J_{\text{Cake}}$ near $\mathcal{D}=4$ obtained in
{\cite{Schroder:2005va}}.

Using Eqs.\thinspace(\ref{eq:j2222}), (\ref{eq:JD}), we obtain
\begin{align}
J_{2}^{(5)}  &  =-\frac{1}{81}\pi^{2}\left(  63\pi^{2}-488-72\zeta_{3}\right)
+\sum_{n=1}^{\infty}T_{2}(n)\,,\label{eq:j2222res}\\
T_{2}(n)  &  =\frac{(-1)^{n}\left(  56n^{3}+80n^{2}+35n+5\right)  \pi
^{2}((n-1)!)^{3}}{18(2n+1)^{2}(3n+1)!}\nonumber\\
&  -\frac{(-1)^{n}2^{-4n-1}\left(  2340n^{3}+744n^{2}+7n-6\right)  \pi
^{4}((2n)!)^{3}}{9n\left(  36n^{2}-1\right)  (n!)^{2}(4n+1)!}\nonumber\\
&  -\frac{16\pi^{2}(2n-1)!(4n-1)!}{27\left(  36n^{2}-1\right)  ^{2}%
(6n+3)!}\nonumber\\
&  \times\left(  11088576n^{6}+7641216n^{5}+691632n^{4}-424512n^{3}%
-79356n^{2}+616n+585\right) \nonumber\\
&  +\frac{2(-1)^{n}\pi^{2}((2n)!)^{5}(3n)!}{27n^{2}\left(  36n^{2}-1\right)
^{2}(n!)^{3}(4n+2)!(6n+1)!}\nonumber\\
&  \times\left(  4489344n^{6}+2794176n^{5}+206352n^{4}-146880n^{3}%
-26032n^{2}-28n+153\right) \nonumber\\
&  -\frac{4\pi^{2}(2n)!(4n)!\left(  H_{2n-1}+2H_{4n-1}-3H_{6n+3}\right)
}{27n^{2}\left(  36n^{2}-1\right)  (6n+3)!}\nonumber\\
&  \times\left(  102672n^{5}+106128n^{4}+33472n^{3}+1960n^{2}-585n-63\right)
\nonumber\\
&  -\frac{(-1)^{n}\pi^{2}((2n)!)^{5}(3n)!\left(  3H_{n-1}-10H_{2n-1}%
-3H_{3n+1}+4H_{4n+2}+6H_{6n+3}\right)  }{54n^{2}\left(  36n^{2}-1\right)
(n!)^{3}(4n+1)!(6n+1)!}\nonumber\\
&  \times\left(  31176n^{4}+15480n^{3}+1594n^{2}-198n-27\right)  \,,\nonumber
\end{align}%
\begin{equation}
J_{2}^{(5)}=0.0516516357945\ldots.
\end{equation}
Here $H_{n}=\sum_{k=1}^{n}k^{-1}$ is a harmonic number. Unfortunately, we have
not been able to express the sums in Eq.\thinspace(\ref{eq:j2222res}) in terms
of $\zeta$-functions and alike. However, the numerical convergence of the
above series is perfect and we obtain from Eqs.\thinspace(\ref{eq:jCakeviaj2}%
), (\ref{eq:j2222res})%
\begin{equation}
J_{\text{Cake}}^{(3+\epsilon)}=\frac{\pi^{4}}{6}-\epsilon\times
58.3184377060\ldots+O\left(  \epsilon^{2}\right)  \,.
\end{equation}

Methods of calculation and values of multiloop vacuum integrals in arbitrary
space-time dimension $\mathcal{D}$ are of independent interest, because these
integrals appear as parts of the amplitudes for various physical processes:
from QCD and QED radiative corrections \cite{Steinhauser:2002rq} to the
thermodynamics of finite temperature QCD-like theories \cite{Kajantie:2003ax}.
In particular, the master integrals in Fig.\,\ref{fig:MIs} enter the basis
intensively used in modern QCD calculations
\cite{Chetyrkin:2006xg,Chetyrkin:2005ia,Faisst:2006sr,Chetyrkin:2006dh}.

\section{Results and Conclusion\label{Sec:Results}}

The perturbative expansion of the form factors in Eq.\thinspace
(\ref{eq:Parametrization}) has the form%
\begin{equation}
f_{i}=\sum_{n=2,4,\ldots}\alpha\left(  Z\alpha\right)  ^{n}f_{i}^{\left(
n\right)  }.
\end{equation}
We have calculated the low-energy expansion of $f_{i}^{\left(  2\right)  }$
and $f_{i}^{\left(  4\right)  }$ up to $O\left(  \lambda^{2}\right)  $, which
corresponds to the expansion of the polarization operator up to $O\left(
\lambda^{4}\right)  $. Using Eq.\thinspace(\ref{eq:POHardSoft}), we obtain in
$(Z\alpha)^{2}$ order%
\begin{align}
f_{1}^{(2)} &  =\frac{7}{2(4!)^{2}}+\frac{1}{6!}\frac{|\mathbf{k}-\mathbf{q}%
|}{m}+\frac{1}{2^{2}(5!)^{2}m^{2}}\left[  \frac{117}{4}\,\omega^{2}%
+49\,\mathbf{k}\cdot\mathbf{q}-\frac{57}{2}\left(  \mathbf{k}^{2}%
+\mathbf{q}^{2}\right)  \right]  ,\label{eq:f12}\\
f_{2}^{(2)} &  =-\frac{73\,}{2^{2}(4!)^{2}}-\frac{11}{6!}\frac{|\mathbf{k}%
-\mathbf{q}|}{m}\nonumber\\
& +\frac{1}{2^{5}(5!)^{2}m^{2}}\left[  -\frac{10923}{2}%
\,\omega^{2}+3095\,\mathbf{k}\cdot\mathbf{q}+1111\left(  \mathbf{k}%
^{2}+\mathbf{q}^{2}\right)  \right]  ,\label{eq:f22}\\
f_{3}^{(2)} &  =-\frac{7}{6!}\frac{|\mathbf{k}-\mathbf{q}|}{m}+\frac
{587}{2^{3}(5!)^{2}}\frac{\left(  \mathbf{k}-\mathbf{q}\right)  ^{2}}{m^{2}%
},\label{eq:f32}\\
f_{4}^{(2)} &  =\frac{573}{2^{4}(5!)^{2}}\frac{\omega^{2}}{m^{2}%
},\label{eq:f42}\\
f_{5}^{(2)} &  =-\frac{4}{6!}\frac{|\mathbf{k}-\mathbf{q}|}{m}+\frac
{1369}{2^{5}(5!)^{2}}\,\frac{\left(  \mathbf{k}-\mathbf{q}\right)  ^{2}}%
{m^{2}}.\label{eq:f52}%
\end{align}
The $O\left(  \lambda^{0}\right)  $ terms of $f_{1,2}^{(2)}$ have been
obtained in Ref.\cite{Costantini:1971cj}. When $\mathbf{k}=\omega=0,$ the
expansion of $f_{1}^{\left(  2\right)  }\left(  q\right)  $ agrees with the
exact result obtained in Ref.\thinspace\cite{Lee:2007}. The next-to-leading terms, proportional to $|\mathbf{k}-\mathbf{q}|$, come from the soft region and exhibit the above-mentioned nonanalytic behavior.

In $(Z\alpha)^{4}$
order we obtain%
\begin{align}
f_{1}^{(4)}= &  \frac{1}{2^{7}}\left(  \frac{2}{\pi^{2}}J_{2}^{\left(
5\right)  }-\frac{4}{3}\,\zeta_{3}+\frac{5}{9}\,\zeta_{2}+\frac{13}%
{18}\right)\label{eq:f14}\\
&  +\frac{1}{2^{10}5\,m^{2}}\left[  \left(  -\frac{109}{2\pi^{2}%
}J_{2}^{\left(  5\right)  }+\frac{475}{3}\,\zeta_{3}-\frac{4111}{30}%
\,\zeta_{2}+\frac{2575}{72}\right)  \omega^{2}\right.  \nonumber\\
&  +\left(  \mathbf{k}^{2}+\mathbf{q}^{2}\right)  \left(  -\frac{6}{\pi^{2}%
}J_{2}^{\left(  5\right)  }-\frac{25}{18}\,\zeta_{2}+\frac{29}{8}\right)
+\left.  \mathbf{k}\cdot\mathbf{q}\left(  -\frac{51}{\pi^{2}}J_{2}^{\left(
5\right)  }+\frac{481}{30}\,\zeta_{2}-\frac{3053}{108}\right)  \right]
,\nonumber\\
f_{2}^{(4)}= &  \frac{1}{2^{7}}\left(  \frac{7}{2\pi^{2}}J_{2}^{(5)}-\frac
{10}{3}\,\zeta_{3}+\frac{1267}{72}\,\zeta_{2}-\frac{3661}{144}\right)\label{eq:f24}
\\
&+\frac{1}{2^{10}5\,m^{2}}\left[  \left(  \frac{473}{4\pi^{2}}J_{2}^{\left(
5\right)  }-145\zeta_{3}+\frac{279829}{720}\,\zeta_{2}-\frac{1233655}%
{2592}\right)  \omega^{2}\right.  \nonumber\\
&  +\left(  \mathbf{k}^{2}+\mathbf{q}^{2}\right)  \left(  \frac{2}{\pi^{2}%
}J_{2}^{\left(  5\right)  }-\frac{95}{2}\,\zeta_{3}-\frac{359}{360}\,\zeta
_{2}+\frac{18769}{288}\right) \nonumber \\
&+\left.  \mathbf{k}\cdot\mathbf{q}\left(
-\frac{235}{2\pi^{2}}J_{2}^{\left(  5\right)  }-\frac{32999}{180}\,\zeta
_{2}+\frac{610}{3}\,\zeta_{3}+\frac{7571}{144}\right)  \right]\nonumber,
\end{align}
\begin{align}
f_{3}^{(4)}= &  \frac{1}{2^{10}5\,}\frac{(\mathbf{k}-\mathbf{q})^{2}}{m^{2}%
}\left(  \frac{87}{2\pi^{2}}J_{2}^{\left(  5\right)  }-\frac{12641}%
{180}\,\zeta_{2}+\frac{24551}{216}\right)  ,\\
f_{4}^{(4)}= &  \frac{1}{2^{10}5\,}\frac{\omega^{2}}{m^{2}}\left(  -\frac
{71}{2\pi^{2}}J_{2}^{\left(  5\right)  }+\frac{25}{3}\,\zeta_{3}+\frac
{2857}{60}\,\zeta_{2}-\frac{111685}{1296}\right)  ,\\
f_{5}^{(4)}= &  \frac{1}{2^{10}5}\frac{(\mathbf{k}-\mathbf{q})^{2}}{m^{2}%
}\left(  \frac{31}{2\pi^{2}}J_{2}^{\left(  5\right)  }-30\zeta_{3}%
+\frac{11639}{180}\,\zeta_{2}-\frac{31319}{432}\right)  .
\end{align}
In $(Z\alpha)^4$ order, the nonanalytic contribution to the form factors is suppressed as $O(\lambda^5)$ and thus is far beyond the accuracy chosen. Note, that the technique used in this paper can be applied without modification to the calculation of the higher terms of the low-energy expansion in $(Z\alpha)^2$ and $(Z\alpha)^4$ orders.

The Coulomb corrections to the form factors $f_{1}$ and $f_{2}$ in order
$\lambda^{0}$ were calculated in Ref.\thinspace\cite{Kirilin2008} numerically.
Although the interaction with the Coulomb field was taken into account
exactly, it turned out that the results can be well fitted by the polynomial
function of $Z\alpha$:%
\begin{align}
f_{1} &  =\frac{7}{2(4!)^{2}}(Z\alpha)^{2}+3.35\cdot10^{-4}(Z\alpha
)^{4}+1.6\cdot10^{-4}(Z\alpha)^{6},\label{eq:fitF1}\\
f_{2} &  =-\frac{73\,}{2^{2}(4!)^{2}}(Z\alpha)^{2}-3.55\cdot10^{-3}%
(Z\alpha)^{4}-2.1\cdot10^{-3}(Z\alpha)^{6}.\label{eq:fitF2}%
\end{align}
In $O\left(  \lambda^{0}\right)  $ order, our results (\ref{eq:f14}),
(\ref{eq:f24}) for the $(Z\alpha)^{4}$-order corrections numerically coincide
with those of Eqs.\thinspace(\ref{eq:fitF1}),\thinspace(\ref{eq:fitF2}) with
an accuracy of a few percent.

\begin{figure}
[ptb]
\begin{center}
\includegraphics[
height=1.6985in,
width=0.8631in
]%
{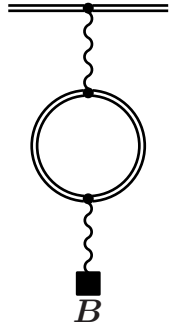}%
\caption{The "magnetic loop" contribution to the bound electron $g$ factor.}%
\label{fig:gBound}%
\end{center}
\end{figure}

As the demonstration of possible applications of our result, let us calculate
the $O\left(  \left(  Z\alpha\right)  ^{7}\right)  $ contribution of the
"magnetic loop" to the $g$ factor of the bound electron, see Fig.\thinspace
\ref{fig:gBound}. The corresponding correction to the $g$ factor of the
electron in $nL_{J}$ state has the form \cite{Lee:2004vb}
\begin{align*}
\frac{\Delta g}{g_{0}} &  =\frac{16}{\pi\left(  2\kappa-1\right)  }\int
\frac{dq}{q}\,f_{1}\left(  q\right)  \,G\left(  q\right)  \,,\\
f_{1}\left(  q\right)   &  =\left.  f_{1}\right\vert _{\mathbf{k}=\omega
=0}\,,\quad G\left(  q\right)  =\int dr\,a\left(  r\right)  b\left(  r\right)
\left(  \sin qr-qr\cos qr\right)  ,\\
g_{0} &  =\frac{2\kappa}{2\kappa+1},\quad\kappa=\left(  J+1/2\right)
\,\mathrm{sign}\left(  L-J\right)  \,,
\end{align*}
where $a$ and $b$ are determined by the form of the bound electron wave
function%
\[
\psi\left(  \mathbf{r}\right)  =\left(
\begin{array}
[c]{c}%
a\left(  r\right)  \Omega_{JLM}\left(  \mathbf{n}\right)  \\
ib\left(  r\right)  \tilde{\Omega}_{JLM}\left(  \mathbf{n}\right)
\end{array}
\right)  .
\]
The characteristic scale of the function $G\left(  q\right)  $ is
$mZ\alpha/n\ll m$, so we have two regions of integration: $q\sim mZ\alpha/n$
and $q\sim m$. In the leading order, only the first region is essential, the
contribution of this region is of the order $O\left(  \left(  Z\alpha\right)
^{5}\right)  $. The leading correction $\sim O\left(  \left(  Z\alpha\right)
^{6}\right)  $ for $L\neq0$ states also comes from the region $q\sim
mZ\alpha/n,$while for $L=0$ the whole interval $mZ\alpha/n\lesssim q\lesssim
m$ is essential. This correction has been found in Ref.\thinspace
\cite{Lee:2004vb}. Note that the integrals in Eq.\thinspace(18) of
Ref.\thinspace\cite{Lee:2004vb} can be taken analytically, and the correction
to $g$ factor up to the order $O\left(  \left(  Z\alpha\right)  ^{6}\right)
$ can be represented as
\begin{align}
\left(  \frac{\Delta g}{g_{0}}\right)  _{\left(  Z\alpha\right)  ^{5}+\left(
Z\alpha\right)  ^{6}} &  =\frac{7\alpha\left(  Z\alpha\right)  ^{5}}%
{288n^{3}J\left(  J+1\right)  \left(  2J+1\right)  }\nonumber\\
&  +\delta_{L=0}\frac{4\alpha\left(  Z\alpha\right)  ^{6}}{135\pi}\frac
{1}{n^{3}}\left[  \log\frac{n}{2Z\alpha}-\frac{641}{240}-H_{n}+\frac{\left(
n+1\right)  \left(  4n-1\right)  }{6n^{2}}\right]  \nonumber\\
&  +\delta_{L\neq0}\frac{2\alpha(Z\alpha)^{6}}{45\pi n^{3}(2L+1)(2\kappa
-1)^{2}}\left(  \frac{3}{L\left(  L+1\right)  }-\frac{1}{n^{2}}\right)  \,.
\end{align}
In order to find the next-to-leading correction, we separate the contributions
of the two regions similar to what has been described above. The details of
this calculation will be presented elsewhere. It turns out that the complete
$O\left(  \left(  Z\alpha\right)  ^{7}\right)  $ result for the correction to
$g$ factor can be expressed via several first term of expansion of the
function $f_{1}$ near $q=0$, namely%
\begin{multline}
\left(  \frac{\Delta g}{g_{0}}\right)  _{\left(  Z\alpha\right)  ^{7}}
=\frac{4\alpha\left(  Z\alpha\right)  ^{7}}{n^{5}J(J+1)(2J+1)}\\
\times\left[
n^{2}\frac{6J\left(  J+1\right)  +1}{(2J+1)^{2}J\left(  J+1\right)  }%
+n\frac{3}{(2J+1)}-2-\frac{1}{2(2\kappa-1)}\right]  f_{1}^{\left(  2\right)
}\left(  0\right)  \\
  -\delta_{L=0}\frac{8\alpha\left(  Z\alpha\right)  ^{7}}{3n^{5}}\left(
1+5n^{2}\right)  m^{2}f_{1}^{\left(  2\right)  \prime\prime}\left(  0\right)
+\frac{4\alpha\left(  Z\alpha\right)  ^{7}}{n^{3}J\left(  J+1\right)  \left(
2J+1\right)  }f_{1}^{\left(  4\right)  }\left(  0\right)  .
\end{multline}
Now, owing to Eq.\thinspace(\ref{eq:f14}), we have the last essential
ingredient to obtain the correction. Using Eqs.\thinspace(\ref{eq:f12}%
),\thinspace(\ref{eq:f14}), we obtain%
\begin{multline}
\left(  \frac{\Delta g}{g_{0}}\right)  _{\left(  Z\alpha\right)  ^{7}}
=\frac{7\alpha\left(  Z\alpha\right)  ^{7}}{288n^{5}J(J+1)(2J+1)}\\
\times\left[n^{2}\frac{6J\left(  J+1\right)  +1}{(2J+1)^{2}J\left(  J+1\right)  }%
+n\frac{3}{(2J+1)}-2-\frac{1}{2(2\kappa-1)}\right]
  +\delta_{L=0}\frac{19\alpha\left(  Z\alpha\right)  ^{7}}{7200n^{5}}\left(
1+5n^{2}\right) \\ +\frac{\alpha\left(  Z\alpha\right)  ^{7}}{32n^{3}J\left(
J+1\right)  \left(  2J+1\right)  }\left(  \frac{2}{\pi^{2}}J_{2}^{\left(
5\right)  }-\frac{4}{3}\,\zeta_{3}+\frac{5}{9}\,\zeta_{2}+\frac{13}%
{18}\right)  .\label{eq:delta_g7}%
\end{multline}
In particular, for $1S_{1/2}$ and $2P_{1/2}$ states we have%
\begin{align}
\left(  \frac{\Delta g}{g_{0}}\right)  _{1S} &  =1.62\times10^{-2}%
\alpha\left(  Z\alpha\right)  ^{5}+9.431\,4\times10^{-3}\alpha\left(
Z\alpha\right)  ^{6}\left(  \ln\frac{1}{2Z\alpha}-2.67\right)\nonumber \\
& +4.1\times
10^{-2}\alpha\left(  Z\alpha\right)  ^{7}\,,\nonumber\\
8\left(  \frac{\Delta g}{g_{0}}\right)  _{2P} &  =1.62\times10^{-2}%
\alpha\left(  Z\alpha\right)  ^{5}+5.894\,6\times10^{-3}\alpha\left(
Z\alpha\right)  ^{6}\nonumber\\
&+3.26\times10^{-2}\alpha\left(  Z\alpha\right)  ^{7}.
\end{align}
The contribution of the $O\left(  \left(  Z\alpha\right)  ^{7}\right)  $ term
is rather essential, e.g., for $Z=6$ (carbon) the ratio of this term to
$O\left(  \left(  Z\alpha\right)  ^{6}\right)  $ term for $1S_{1/2}$ state is
$-0.81$. The last term in Eq.\thinspace(\ref{eq:delta_g7}) corresponds to the
contribution of the electron loop with four Coulomb exchanges. It is
interesting to compare the magnitude of this term with that of the first two
terms. As it was claimed in Ref.\thinspace\cite{Lee:2004vb} this term appears
to be numerically small. E.g., for the ground state, the contribution of the
last term is only $2.2$ percent.

\appendix

\section{Appendix}

The explicit form of the four simple master integrals from Fig.\thinspace
\ref{fig:MIs} reads:%
\begin{align}
J_{\text{Infinity}}^{\left(  \mathcal{D}\right)  } & \overset{def}{\equiv
}J_{0101111100}^{(\mathcal{D})} \nonumber \\
& =\frac{\Gamma(6-2\mathcal{D})\Gamma\left(
5-3\mathcal{D}/2\right)  ^{2}\Gamma\left(  2-\mathcal{D}/2\right)  ^{2}%
\Gamma\left(  \mathcal{D}/2-1\right)  ^{4}\Gamma\left(  3\mathcal{D}%
/2-4\right)  }{\Gamma(10-3\mathcal{D})\Gamma(\mathcal{D}-2)^{2}\Gamma\left(
\mathcal{D}/2\right)  },\nonumber\\
J_{\text{Melon}}^{\left(  \mathcal{D}\right)  } & \overset{def}{\equiv
}J_{0011110100}^{(\mathcal{D})}   =\frac{\Gamma(5-2\mathcal{D})\Gamma\left(
4-3\mathcal{D}/2\right)  ^{2}\Gamma(3-\mathcal{D})\Gamma\left(  \mathcal{D}%
/2-1\right)  ^{3}}{\Gamma(8-3\mathcal{D})\Gamma\left(  \mathcal{D}/2\right)
},\nonumber\\
J_{\text{Tumbler}}^{\left(  \mathcal{D}\right)  } & \overset{def}{\equiv
}J_{0111100100}^{(\mathcal{D})}   =\frac{\Gamma\left(  4-3\mathcal{D}%
/2\right)  \Gamma(3-\mathcal{D})^{2}\Gamma\left(  1-\mathcal{D}/2\right)
\Gamma\left(  2-\mathcal{D}/2\right)  \Gamma\left(  \mathcal{D}/2-1\right)
^{2}}{\Gamma(6-2\mathcal{D})\Gamma\left(  \mathcal{D}/2\right)  },\nonumber\\
J_{\text{Clover}}^{\left(  \mathcal{D}\right)  } & \overset{def}{\equiv
}J_{1111000000}^{(\mathcal{D})}   =\Gamma\left(  1-\mathcal{D}/2\right)
^{4}. \label{eq:MIsimple}%
\end{align}
The coefficients in Eq.\thinspace(\ref{eq:j2222}) are%
\begin{align}
a_{\text{Cake}}  &  =\frac{(\epsilon+1)^{3}\left(  9\epsilon^{2}%
+3\epsilon-4\right)  }{48(3\epsilon+2)},\nonumber\\
\frac{a_{\text{Clover}}}{\epsilon^{3}}  &  =\frac{(\epsilon+1)(\epsilon+3)^{3}\left(
6\epsilon^{3}+5\epsilon^{2}+3\epsilon+2\right)  }{384\epsilon^{3}%
(3\epsilon+2)},\nonumber\\
a_{\text{Infinity}}  &  =-\left(  \frac{9\epsilon^{3}}{16}-\frac{3\epsilon
^{2}}{8}-\frac{51\epsilon}{16}-\frac{5}{3}\right)  \frac{(\epsilon
+1)_{2}(2\epsilon+2)_{3}}{(3\epsilon+1)_{4}},\nonumber\\
\frac{a_{\text{Tumbler}}}{\epsilon^{2}}  &  =-\left(  4\epsilon^{6}+\frac
{73\epsilon^{5}}{6}+16\epsilon^{4}+\frac{967\epsilon^{3}}{54}+\frac
{397\epsilon^{2}}{27}+\frac{17\epsilon}{3}+\frac{2}{3}\right)  \frac
{(\epsilon+1)\left(  3\epsilon/2+3/2\right)  _{4}}{(2\epsilon)_{4}%
(3\epsilon)_{3}},\nonumber\\
\frac{a_{\text{Melon}}}{\epsilon^{2}}  &  =\left(  \frac{9\epsilon^{5}}{16}%
-\frac{3\epsilon^{4}}{4}-\frac{117\epsilon^{3}}{16}-\frac{305\epsilon^{2}}%
{24}-\frac{51\epsilon}{8}-\frac{3}{4}\right)  \frac{\left(  3\epsilon
/2+5/2\right)  _{2}(2\epsilon+2)_{4}}{\epsilon(\epsilon+2)(3\epsilon)_{5}},
\end{align}
Here $x_{n}=\Gamma\left(  x+n\right)  /\Gamma\left(  x\right)  =x\left(
x+1\right)  \ldots\left(  x+n-1\right)  $. The coefficients in Eq.\thinspace
(\ref{eq:recurrence}) are%
\begin{align}
\epsilon b_{\text{Cake}}  &  =-\frac{\epsilon(\epsilon+1)^{3}(\epsilon
+2)(\epsilon+3)}{48(3\epsilon-2)(3\epsilon+2)},\nonumber\\
\epsilon^{-2}b_{\text{Clover}}  &  =\frac{(\epsilon+1)(\epsilon+3)^{4}\left(
14\epsilon^{3}+40\epsilon^{2}+35\epsilon+10\right)  }{384\epsilon
^{2}(3\epsilon-2)(3\epsilon+2)},\nonumber\\
b_{\text{Infinity}}  &  =-\left(  3-\frac{7\epsilon}{4}-93\epsilon^{2}%
-\frac{585\epsilon^{3}}{4}\right)  \frac{(\epsilon+2)\epsilon_{4}%
(2\epsilon+2)_{3}}{4\,(3\epsilon-2)_{7}},\nonumber\\
\epsilon^{-1}b_{\text{Tumbler}}  &  =\left(  7+\frac{65\epsilon}{2}%
-\frac{490\epsilon^{2}}{9}-\frac{4184\epsilon^{3}}{9}-737\epsilon^{4}%
-\frac{713\epsilon^{5}}{2}\right)  \frac{(\epsilon+1)_{3}\left(
3\epsilon/2+5/2\right)  _{3}}{64\,\left(  \epsilon+1/2\right)  _{2}%
(3\epsilon-2)_{5}},\nonumber\\
\epsilon^{-1}b_{\text{Melon}}  &  =\left(  3+11\epsilon-\frac{797\epsilon^{2}%
}{18}-215\epsilon^{3}-\frac{433\epsilon^{4}}{2}\right)  \frac{3\,\left(
3\epsilon/2+5/2\right)  _{2}(2\epsilon+2)_{5}}{16\,(3\epsilon-2)_{7}}.
\end{align}

\end{document}